\documentclass[epjCONF]{svjour}

\usepackage{graphics,graphicx}
\usepackage[varg]{txfonts} 
\usepackage[latin1]{inputenc}

\usepackage{latexsym,amssymb,mathrsfs,bm}

\newcommand{\be}{\begin{equation}}
\newcommand{\ee}{\end{equation}}
\newcommand{\bea}{\begin{eqnarray}}
\newcommand{\eea}{\end{eqnarray}}
\newcommand{\bse}{\begin{subequations}}
\newcommand{\ese}{\end{subequations}}

\newcommand{\tr}{\textrm{tr}}
\newcommand{\tn}[1]{{\textnormal{#1}}}
\newcommand{\ts}[1]{{\textnormal{\scriptsize #1}}}
\newcommand{\grdstate}[1]{\left\langle #1 \right\rangle }
\def\p{{\bf p}}
\def\k{{\bf k}}

\newcommand{\slk}{\mbox{\,\slash \hspace{-0.5em}$k$}}

\newcommand{\vecbm}[1]{{\bm{#1}}}

\session-title{Hot and Cold Baryonic Matter -- HCBM 2010}

\begin{document}

\title{The effect of the Polyakov loop on the chiral phase transition}
\author{G. Mark{\'o}\inst{1}\fnmsep\thanks{\email{smatkovics@hotmail.com}}
  \and Zs. Sz{\'e}p \inst{2}\fnmsep\thanks{\email{szepzs@achilles.elte.hu}\,.
Speaker. On leave from Statistical and Biological Physics Research Group
of the Hungarian Academy of Sciences, H-1117 Budapest, Hungary.}}

\institute{Department of Atomic Physics, E{\"o}tv{\"o}s University, H-1117 Budapest, Hungary \and Centre de Physique Th{\'e}orique, Ecole Polytechnique,
CNRS, 91128 Palaiseau Cedex, France}

\abstract{The Polyakov loop is included in the $SU(2)_L\times SU(2)_R$ 
chiral quark-meson model by considering the propagation of the constituent
quarks, coupled to the $(\sigma,\vecbm{\pi})$ meson multiplet, on the
homogeneous background of a temporal gauge field, diagonal in color
space.  The model is solved at finite temperature and quark baryon
chemical potential both in the chiral limit and for the physical value
of the pion mass by using an expansion in the number of flavors $N_f.$
Keeping the fermion propagator at its tree-level, a resummation
on the pion propagator is constructed which resums infinitely many
orders in $1/N_f,$ where ${\cal O}(1/N_f)$ represents the order at
which the fermions start to contribute in the pion propagator. The
influence of the Polyakov loop on the tricritical or the critical
point in the $\mu_q-T$ phase diagram is studied for various forms of
the Polyakov loop potential.} 

\maketitle

\section{Introduction}
\label{intro}

The low-energy effective models of the QCD, such as the
Nambu--Jona-Lasinio (NJL) model and the chiral quark-meson model (QM)
are based on the global chiral symmetry of the QCD. They are very
useful to qualitatively understand many aspects related to the
spontaneous breaking of the chiral symmetry and its restoration at
finite temperature and density. However, the absence of gluonic
effective degrees of freedom and the lack of color clustering
alter the reliability of the quantitative thermodynamic predictions of
these models, such as the equation of state or the location of the
critical end point (CEP) in the $\mu_q-T$ phase diagram.

Some information on the quark confinement can be incorporated in the
effective models through an effective degree of freedom, the Polyakov
loop, which is a good order parameter for the deconfinement phase
transition in the absence of dynamical quarks. The coupling of the
Polyakov loop to the chiral effective models mimics the effect of
confinement by statistically suppressing at low temperature the
contribution of one- and two-quark states relative to the three-quark
states. This feature makes the Polyakov-loop extended effective models
more appropriate for the description of the low-temperature phase and
for quantitative comparison with the thermodynamic observables on the
lattice \cite{weise08,schaefer07,schaefer10}. Better agreement is
expected up to $T\simeq (1.5-2) T_c$ above which the transverse
gluonic degrees of freedom dominate in thermodynamic quantities, such
as the pressure, over the longitudinal ones represented by the
Polyakov loop.

Some solutions of the PQM model appearing in the literature completely
disregard quantum effects. The effect of including the quantum
fluctuation in the PQM model was recently studied in
\cite{nakano10,skokov10,herbst10} using functional renormalization
group methods and also in \cite{marko10}, where it was shown that the
inclusion of the fluctuations has a significant effect on the location
of the CEP, which is pushed to higher values of $\mu_q.$ In this
contribution we review the results on the $\mu_q-T$ phase diagram
obtained in \cite{marko10} as a result of including different forms of
the effective Polyakov loop potential, as compared to those previously
obtained in \cite{toni04} in the chiral limit of the two flavor QM
using the resummation of the perturbative series provided by the
large-$N_f$ approximation. Starting with the $\Phi$-derivable
formalism, the approximations done to parametrize and solve the model
are also discussed.

\section{The model in the large-$N_f$ approximation}

The ingredients for the Polyakov loop extended quark-me\-son model
(PQM) are the $(\sigma,\vecbm{\pi})$ meson multiplet and the 
constituent quark fields $\psi$ coupled to them. These latter
propagate on the homogeneous background of a temporal gauge field. In
order to be able to use a large-$N_f$ expansion we consider $N_f$
constituent quarks and correspondingly $N-1$ pions ($\sqrt{N}=N_f$).
Performing some rescaling with $N_f$ (see \cite{marko10,toni04} for
details), which assures the finiteness of the tree-level constituent
quark mass $m_q=g v$ as $N_f\to\infty,$ the Lagrangian of the model
reads after separating the vacuum expectation value $v N_f$ of the
$\sigma$ field as~: \bea {\cal L}&=& -N\left[\frac{\lambda}{24}
  v^4+\frac{1}{2}m^2v^2 -h v\right]- \sqrt{N}\left[\frac{\lambda}{6}
  v^3+m^2 v - h\right]\sigma
\nonumber\\ &+&\frac{1}{2}\bigl[(\partial\sigma)^2 +
  (\partial\vecbm{\pi})^2 \bigr] -\frac{1}{2}m^2_{\sigma 0}\sigma^2
-\frac{1}{2}m^2_{\pi 0}\vecbm{\pi}^2\nonumber\\ &-&\frac{\lambda
  v}{6\sqrt{N}}\sigma \rho^2- \frac{\lambda}{24N}\rho^4 +\bar
\psi\big[(i\partial^\mu+\delta^{\mu0}A_0)\gamma_\mu-m_q\big]\psi
\nonumber\\ &-&\frac{g}{\sqrt{N}}\left[\bar\psi \left(\sigma
  +i\sqrt{2N_f}\gamma_5T^a\pi^a\right)\right]\psi,
\label{Eq:Lagrange_field}
\eea  
where $\rho^2=\sigma^2+\vecbm{\pi}^2$ and the tree-level sigma and pion
masses are $m^2_{\sigma 0}=m^2+\lambda v^2/2$ and 
$m^2_{\pi 0}=m^2+\lambda v^2/6.$

\subsection{The grand potential in the $\Phi$-derivable approximation}

At finite temperature $T=1/\beta$, after analytical continuation to
imaginary time $t\to i\tau$, $A_0 \to iA_4$, the grand partition
function $Z$ and the grand potential $\Omega(T,\mu_B)$ of the
spatially uniform system defined by (\ref{Eq:Lagrange_field}) are
defined as
\be
Z=\tr\big\{
\exp\big[-\beta\big( H_0(A_4)+H_\ts{int}-\mu_B Q_B\big)\big]
\big\}=e^{-\beta \Omega},
\label{Eq:Z_def}
\ee
where $\mu_B$ is the baryon chemical potential. $H_\ts{int}$ is the 
interacting part of the Hamiltonian constructed from (\ref{Eq:Lagrange_field}).
The $A_4$-dependent free Hamiltonian $H_0(A_4)$ for two quark flavors 
$u$ and $d$ reads
\be
H_0(A_4)=H_0+\int d^3x \left[i u^\dagger_i(x) A_{4}^{ij} u_j(x) + 
i d^\dagger_i(x) A_{4}^{ij} d_j(x)\right],
\ee
where $H_0$ is the quadratic part of the Hamiltonian at vanishing 
and $i,j$ denotes color indices. In the so-called Polya\-kov gauge 
$A_4$ is diagonal in color space: $A_4=\tn{diag}(\phi_+,\phi_-,$ 
$-(\phi_++\phi_-)).$ The conserved baryon charge $Q_B$ appearing in 
(\ref{Eq:Z_def}) can be expressed in terms of the particle number 
operators as $Q_B=\frac{1}{3}\sum\limits_{i=1}^3 N_{q,i},$ with
$N_\ts{q,i}=N_{u,i}+N_{d,i}-N_{\bar u,i}-N_{\bar d,i}$ and {\it e.g. } 
$N_{u,i}=\int d^3 x \big(u_i^\dagger u_i+d_i^\dagger d_i\big).$
Then, combining $H_0(A_4)$ and $\mu_B Q_B,$ the effect of 
fermions propagating on the constant background $A_4$, diagonal in 
the color space is like having imaginary chemical potential for color.
Following Ref.~\cite{korthals_altes00}, one introduces 
color-dependent chemical potential for fermions
\be
\mu_{1,2}=\mu_q-i\phi_\pm,\quad 
\mu_3=\mu_q+i(\phi_++\phi_-),
\label{Eq:c-dep_mu}
\ee
where $\mu_q=\mu_B/3$ is the quark baryon chemical potential.
Then, introducing the notation 
${\cal H}=H_0- \sum\limits_{i=1}^3\mu_i N_{q,i},$ 
one can write $Z$ as a path integral over the fields, generically 
denoted by~$\Psi$ 
\be
Z=e^{-\beta \Omega_0}
\frac{\displaystyle
\int\big[{\cal D}\Psi\big] \bigg\{
e^{-\beta {\cal H}} 
{\cal P} \exp\Big[-\int_0^\beta d\tau H_\ts{int}(\tau)\Big] \bigg\}
}
{\displaystyle \int\big[{\cal D}\Psi\big] 
e^{-\beta {\cal H}}},
\ee
where $\Omega_0$ is the grand potential of the
unperturbed system with fermions having color-dependent chemical potential 
\be
e^{-\beta \Omega_0}=
\int\big[{\cal D}\Psi\big] e^{-\beta {\cal H}}.
\ee

In the $\Phi$-derivable approximation of
Ref.~\cite{ward60}, which is also
called two-particle irreducible (2PI) approximation, the grand
potential $\Omega\equiv\Omega[G_\pi,G_\sigma,G,v,\Phi,\bar\Phi]$
is a functional of the full propagators and field
expectation values of the form~:
\bea
\beta\Omega&=&U(\Phi,\bar\Phi)+
\frac{N}{2} m^2 v^2+N\frac{\lambda}{24} v^4 - N h v
\nonumber\\
&-&(N-1)\frac{i}{2}\int_k 
\left[\ln G_\pi^{-1}(k)+D^{-1}_\pi(k) G_\pi(k)\right]
\nonumber\\
&-&
\frac{i}{2}\int_k\left[
\ln G_\sigma^{-1}(k)+D^{-1}_\sigma(k) G_\sigma(k)\right]
\nonumber\\
&+&\sqrt{N} i\, \tr_{D,c}\int_k 
\left[\ln G^{-1}(k)+D^{-1}(k) G(k)\right]
+\Gamma_\ts{2PI}\, ,
\label{Eq:Omega_grand_pot}
\eea
where the trace is taken in Dirac and color space. 
The tree-level propagators of the pion, sigma, 
and constituent quark fields are 
\be
i D_{\pi/\sigma}^{-1}(k)=k^2-m^2_{\pi/\sigma0},\qquad
i D^{-1}(k)=\slk-m_q,\qquad
\ee
while $G_\pi,$ $G_\sigma,$ and $G$ are the respective full propagators
in terms of which the set of 2PI skeleton diagrams denoted by
$\Gamma_\ts{2PI}\equiv\Gamma_\ts{2PI}[G_\pi,G_\sigma,G,v,\Phi,\bar\Phi]$
is constructed. To ${\cal O}(1/\sqrt{N})$ accuracy this is given by
\bea
\Gamma_\ts{2PI} &=&
N\frac{\lambda}{24}\left(\int_k G_\pi(k)\right)^2
+\frac{\lambda}{12} \int_k G_\pi(k) \int_p G_\sigma(p)
\nonumber\\
&-&\frac{\lambda}{12}i\int_k \Pi(k) 
-\frac{i}{2}\int_k\ln\bigg(1-\frac{\lambda}{6}\Pi(k)\bigg)
\nonumber\\
&-&\frac{\lambda}{6} v^2\int_k G_\sigma(k)
+\frac{\lambda}{6} v^2\int_k \frac{G_\sigma(k)}{1-\lambda \Pi(k)/6}
\nonumber\\
&-&\sqrt{N}\frac{g^2}{2}i\tr_{D,c}\int_k\int_p\gamma_5 G(k)\gamma_5
G(k+p)G_\pi(p)
\nonumber \\
&+&\frac{g^2}{2\sqrt{N}}i\tr_{D,c}\int_k\int_p G(k) G(k+p) G_\sigma(p)\,, \ \ \
\label{Eq:Omega_2PI}
\eea
where the notation $\displaystyle \Pi(k)=-i\int_p G_\pi(p) G_\pi(k+p)$
was introduced. The mesonic part of $\Gamma_\ts{2PI}$ contains the 2PI
diagrams of the $O(N)$ model as given in Eq.~(2.13) and Figs.~2 and 4 of
\cite{dominici93} and also in Eq.~(48) and Fig.~2 of \cite{fejos09}.
Notice that the contribution of the fermions goes with
fractional powers of $N$ and intercalates between the leading order (LO)
and next-to-leading order (NLO) contributions of the pions, which go
with integer powers of $N.$ 

\subsection{The mean-field Polyakov-loop potential\label{ss:PEP}}

$U(\Phi,\bar\Phi)$ appearing in (\ref{Eq:Omega_grand_pot}) represents
one of the different versions of effective Polyakov-loop potentials
defined in the literature for $\mu_q=0,$ where
$|\Phi|=|\bar\Phi|.$ The simplest effective potential is of a Landau
type, constructed with terms consistent with the $\mathbb{Z}_3$
symmetry \cite{pisarski00}:
\be
\beta^4\,{\cal U}_\ts{poly}(\Phi,\bar\Phi)=
-\frac{b_2(T)}{2}\Phi\bar\Phi -\frac{b_3}{6}(\Phi^3 + \bar\Phi^3)
+\frac{b_4}{4}(\Phi\bar\Phi)^2\, ,
\label{Eq:P_eff_pot_poly}
\ee  
where the temperature-dependent coefficient which makes
spontaneous symmetry breaking possible is 
\be
b_2(T)=a_0 + a_1\left(\frac{T_0}{T}\right) +a_2\left(\frac{T_0}{T}\right)^2 
+a_3 \left(\frac{T_0}{T}\right)^3\,.
\ee
$T_0$ is the transition temperature of the confinement/de\-con\-fine\-ment
phase transition, in the pure gauge theory $T_0=270$~MeV.  The
parameters $a_i, i=0,\dots, 3$ and $b_3, b_4$ determined in
\cite{ratti06} reproduce the data measured in pure $SU(3)$ lattice
gauge theory for pressure, and entropy and energy densities.  When
using this potential in either the PNJL or the PQM models the minimum
of the resulting thermodynamic potential is at $\Phi>1$ for
$T\to\infty,$ which is not in accordance with the value coming from
the definition 
$\Phi(\vec x)=\grdstate{\tr_c L(\vec x)}/N_c$ with  
$L(\vec x)={\cal P} \exp\left[i\int_0^\beta d\tau A_4(\tau,\vec x) \right].$ 

An effective potential for the Polyakov loop inspired by a
strong-coupling expansion of the lattice QCD action was derived in
\cite{fukushima04b}. Using the part coming from the
$SU(3)$ Haar measure of group integration an effective potential was
constructed in \cite{ratti07}
\bea
\beta^4\,{\cal U}_{\ts{log}}(\Phi,\bar\Phi)&=& -\frac{1}{2}a(T) \Phi\bar \Phi 
+ b(T) \ln \big[1-6 \Phi\bar\Phi 
\nonumber\\
&+&4(\Phi^{3}+\bar\Phi^{3})
  - 3 (\Phi\bar\Phi)^{2}\big]\, ,
\label{Eq:P_eff_pot_log}
\eea
with the temperature-dependent coefficients
\be
a(T)=a_0+a_1 \left(\frac{T_0}{T}\right)+a_2 \left(\frac{T_0}{T}\right)^2,
\quad
b(T)=b_3\left(\frac{T_0}{T}\right)^3\ .
\ee
The parameters $a_i,i=0,1,2$ and $b_3$ reproduce the thermodynamic
quantities in the pure $SU(3)$ gauge theory measured on the
lattice. Since the logarithm in ${\cal U}_{\ts{log}}(\Phi,\bar\Phi)$
diverges as $\Phi,\bar\Phi\to 1$ the use of this effective potential
guarantees that $\Phi,\bar\Phi\to 1$ for $T\to \infty.$

A third potential is the one determined in 
Refs.~\cite{fukushima04b}:
\bea
\beta\, {\cal U}_{\ts{Fuku}}(\Phi,\bar\Phi) &=& 
-b \left[
54 e^{-a/T} \Phi \bar\Phi +\ln\left(1 - 6 \Phi \bar\Phi\right.\right.
\nonumber\\ 
&+&\left.\left. 4 (\Phi^3 + \bar\Phi^3)
- 3 (\Phi \bar\Phi)^2 \right)
\right],
\label{Eq:P_eff_pot_Fuku}
\eea
where the temperature of the deconfinement phase
transition in pure gauge theory is controlled by $a$, 
while $b$ controls the weight of 
gluonic effects in the transition. In this case, the parameters
$a=664$~MeV and $b=(196.2 \tn{MeV})^3$ are obtained from the
requirement of having a first order transition at about $T=270$~MeV
\cite{fukushima08,schaefer07}. 

It was shown in \cite{fukushima08} that there is little difference in
the pressure calculated from the three effective potentials for the
Polyakov loop in their validity region up to $T\simeq (1.5-2) T_c.$
The presence of dynamical quarks influences an effective treatment
based on the Polyakov loop which in this case is not an exact order
parameter.  Defining effective Polyakov-loop potentials for
nonvanishing chemical potential when $|\Phi|\ne|\bar\Phi|$ is somewhat
ambiguous \cite{schaefer07}. Nevertheless, at $\mu_q\ne 0$ we use the
$\mathbb{Z}_3$ symmetric Polyakov-loop potentials given above and
include the effect of the dynamical quarks along the lines of
\cite{schaefer07}, where using renormalization group arguments the
$N_f$ and $\mu_q$ dependence of the $T_0$ parameter of the
Polyakov-loop effective potential was parametrized as
$T_0(\mu_q,N_f)=T_\tau \exp(-1/(\alpha_0 b(\mu_q)))$.  The parameters
were chosen to have $T_0(\mu_q=0,N_f=2)=208.64$~MeV. When using the
Polyakov loop effective potential given in (\ref{Eq:P_eff_pot_poly})
and (\ref{Eq:P_eff_pot_log}) we will consider in addition to
$T_0=270$~MeV two more cases, one with a constant value $T_0=208$~MeV
and the other with the above-mentioned $\mu_q$-dependent $T_0$ taken
at $N_f=2.$

\subsection{The propagator and field equations and the approximations made to solve them\label{ss:approx}}
We are interested in the equations for the two-point functions and the 
field equations, which are given by the stationary conditions
\be
\frac{\delta \Omega}{\delta G}=\frac{\delta \Omega}{\delta G_\pi}=
\frac{\delta \Omega}{\delta G_\sigma}=\frac{\delta \Omega}{\delta v}=
\frac{\delta \Omega}{\delta \Phi}=\frac{\delta \Omega}{\delta \bar\Phi}=0.
\label{Eq:stac_cond}
\ee 
In each of these equations the contribution of the fermions is kept
only at LO in the large-$N_f$ expansion. This contribution is 
${\cal O}(\sqrt{N})$ in the field equations of $\Phi$ and $\bar\Phi,$
${\cal O}(1)$ in the equation for the fermion propagator $G,$ and
${\cal O}(1/\sqrt{N})$ in the remaining equations, that is the field
equation of $v$, and the equations of $G_\pi$ and $G_\sigma.$

The second line of (\ref{Eq:Omega_2PI}) does not contribute to any of
the equations at the order of interest, and the second term on
the right hand side of (\ref{Eq:Omega_2PI}) contributes
only in the equation of the sigma propagator 
\bea
i G_\sigma^{-1}(p)=i D_\sigma^{-1}(p)+\frac{\lambda v^2}{3}-
\frac{\lambda}{6}\int_k G_\pi(k)\qquad\qquad\quad
\nonumber\\
-\frac{\lambda v^2}{3} \frac{1}{1-\lambda \Pi(p)/6}
-\frac{i g^2}{\sqrt{N}}\tr_{D,c}\int_k G(k) G(k+p)\,.\ \ 
\label{Eq:Gs}
\eea
In fact, $G_\sigma$ will not appear in any of the remaining five
equations. Nevertheless,
it plays an important role in the parametrization of the model, 
as discussed in Sec.~\ref{ss:param}.

In the imaginary time formalism of the finite temperature field theory,
which we use for calculation, the four-momentum is $k=(i\omega_n,\k),$
where the Matsubara frequencies are $\omega_n=2\pi n T$ for bosons,
while for fermions they depend also on the color due to the color-dependent 
chemical potential $\mu_i$ introduced in (\ref{Eq:c-dep_mu}) and are given by
$\omega_n=(2 n+1)\pi T-i \mu_i.$ The meaning of the integration symbol 
encountered so far is then
\be
\int_k = i T\sum_{n} \int_{\k}\equiv i T\sum_{n}\int\frac{d^3 \k}{(2\pi)^3}.
\label{Eq:sum_int_def}
\ee
The dependence on $\Phi$ and $\bar\Phi$ of the fermionic trace-log
term in the grand potential $\Omega$ and of the quark-pion setting-sun
in $\Gamma_\ts{2PI}$ results from the fact that, after performing the
Matsubara sum, the color trace can be expressed in closed form in terms
of the mean-field ($\vec{x}$-independent) Polyakov loop $\Phi$ and its
conjugate $\bar \Phi.$ The big difference is that while in case of the
trace-log the result can be expressed in terms of a modified
Fermi-Dirac distribution function
\be
\tilde f_\Phi^+(E)=
\frac{\left(\bar\Phi +2\Phi e^{-\beta(E-\mu_q)}\right) e^{-\beta(E-\mu_q)}
      + e^{-3\beta(E-\mu_q)} }
{1 + 3\left( \bar\Phi + \Phi e^{-\beta(E-\mu_q)} \right) e^{-\beta(E-\mu_q)}
      + e^{-3\beta(E-\mu_q)}}, 
\label{Eq:F_P+}
\ee
and a similar expression for $\tilde f_\Phi^-(E),$ but with 
$\Phi\leftrightarrow\bar\Phi$ and $\mu_q\leftrightarrow-\mu_q,$
in case of the setting-sun integral the result does not allow for an
interpretation in terms of distribution functions, as shown in the
Appendix of \cite{marko10} between Eqs.~(A35) and (A37).

Due to the complexity of the problem, the set of coupled equations
coming from (\ref{Eq:stac_cond}) is only solved using some
approximations described below. 

1.~As a first approximation we disregard the self-con\-sis\-tent equation for 
the fermions arising from $\delta\Omega/\delta G=0,$ that is
\be
i G^{-1}(k)=i D^{-1}(k)-i g^2\int_p \gamma_5 G(p) \gamma_5 G_\pi(p-k),
\label{Eq:G}
\ee 
and simply use  the tree-level fermion propagator $D(k)$ in the
remaining five equations. Within this approximation the field equation of
$v$ and the pion propagator simplify considerably. The contribution of
the last but one term of (\ref{Eq:Omega_2PI}) to the pion propagator 
breaks up upon working out the Dirac structure into the linear
combination of a fermionic tadpole $\tilde T(m_q)$ and a bubble
integral $\tilde I(p;m_q)$. Introducing the propagator
\be
D_0(k)=\frac{i}{k^2-m_q^2},
\label{Eq:D0_prop}
\ee
these integrals are defined as
\bea
\label{Eq:T_q_def}
\tilde T(m_q)&=&\frac{1}{N_c}\sum_{i=1}^{N_c}\int_k D_0(k),\\
\tilde I(p;m_q)&=&\frac{1}{N_c}\sum_{i=1}^{N_c}
\left[-i\int_q D_0(q) D_0(q+p)\right].
\label{Eq:I_q_def}
\eea
In terms of these integrals given between Eqs.~(A11) and (A18) of 
the Appendix of \cite{marko10}, where the sum over the color indices
is explicitly done, one obtains
\bea
h&=&v\left[m^2+\frac{\lambda}{6}\left(v^2+\int_k G_\pi(k)\right)
-\frac{4 g^2 N_c}{\sqrt{N}}\tilde T(m_q) \right],\quad\ 
\label{Eq:EoS}
\\
i G_\pi^{-1}(k)&=&k^2-m^2-\frac{\lambda}{6}\left(v^2+\int_k G_\pi(k)\right)
+ \frac{4g^2 N_c}{\sqrt{N}} \tilde T(m_q) 
\nonumber\\ 
&-& \frac{2 g^2 N_c}{\sqrt{N}} k^2 \tilde I(p;m_q)\, .
\label{Eq:Gp}
\eea  
By making use of the field equation for $v$ in the equation for the
pion propagator one finds that $i G_\pi^{-1}(k=0)=-h/v,$ which means 
that the Goldstone theorem is fulfilled. Unfortunately, it turns out 
that this is only accidental, because the Ward identity relating the
inverse fermion propagator and the proper vertex
$\Gamma_{\pi^a\psi\bar\psi}=\delta^3 \Gamma/\delta \bar\psi\delta \psi \delta \pi^a$ (see {\it e.g.} Eq.~(13.102) of \cite{zinn-justin})
\be
-\frac{i}{2} T_a\big\{\gamma_5,i G^{-1}(p)\big\}=v\sqrt{\frac{N}{2 N_f}}
\Gamma_{\pi^a\psi\bar\psi}(0,p,-p),
\label{Eq:WI}
\ee
is satisfied only with tree-level propagators and vertices.  The
relation above is violated at any order of the perturbation theory in
the large-$N_f$ approximation, for in view of (\ref{Eq:G}) the
corrections to the inverse tree-level fermion propagator are of 
${\cal O}(1),$ while the corrections to the tree-level $\pi-\psi-\bar\psi$
vertex are suppressed by $1/N.$ The feature is probably a shortcoming 
of our way of implementing the large-$N_f$ scaling in 
(\ref{Eq:Lagrange_field}), and we could note find a way to overcome it.

2.~A further approximation concerns the self-consistent pion
propagator (\ref{Eq:Gp}). In \cite{marko10} several approximations for 
$G_{\pi}$ were discussed, here we consider only two of them.

In the chiral limit a local approximation is obtained by parametrizing
the pion propagator as
\be
G_{\pi,l}(p)=\frac{i}{p^2-M^2},
\label{Eq:Gp_local}
\ee
which is then used in all of the equations. $M^2$ is determined from 
$M^2=-i G^{-1}_{\pi,l}(p=0),$ which gives the gap-equation
\be
M^2=m^2+\frac{\lambda}{6}\left(v^2+T_F(M)\right)
-\frac{4g^2 N_c}{\sqrt{N}} \tilde T_F(m_q).
\label{Eq:gap_p0}
\ee
The subscript $F$ denotes the finite part of the cutoff regularized
integrals defined in Eqs.~(\ref{Eq:T_q_def}) and (\ref{Eq:I_q_def}).
In view of the field equation (\ref{Eq:EoS}) in the chiral limit
$h=0,$ one has $M^2=0.$ We note that due to their self-consistent
nature, when (\ref{Eq:gap_p0}) is solved, a series containing all
orders of $1/\sqrt{N}$ is in fact resummed.

In the case of a physical pion mass, in addition to the local
approximation (\ref{Eq:Gp_local}) and (\ref{Eq:gap_p0}) for the pion
propagator, a nonlocal approximation is derived using an expansion to
${\cal O}(1/\sqrt{N})$ in the expression of the pion propagator
(\ref{Eq:Gp}) obtained after exploiting the field equation of $v$
(\ref{Eq:EoS}):
\bea
G_\pi(p)
&=&\frac{i}{p^2-\frac{h}{v}}+\frac{2 g^2 N_c}{\sqrt{N}} 
\frac{i p^2 \tilde I_F(p;m_q)}{\left(p^2-\frac{h}{v}\right)^2}
+{\cal O}\left(\frac{1}{N}\right).
\label{Eq:Gp_third}
\eea  
With this form of the pion propagator the field equation for $v$ reads
\bea
m^2+\frac{\lambda}{6}\left(v^2+T_F(M)\right)
+\frac{2 g^2 N_c}{\sqrt{N}} J_F(M,m_q)
\nonumber\\
-\frac{4 g^2 N_c}{\sqrt{N}} \tilde T_F(m_q)=\frac{h}{v},
\label{Eq:EoS_third}
\eea
where in this case $M^2=h/v$ and we have introduced the integral
\be
J(M,m_q)=-i\int_p G^2_{\pi,l}(p) p^2 \tilde I_F(p;m_q).
\label{Eq:J_def}
\ee
Solving this equation for $v$ shows that this approximation also resums 
infinitely many orders in $1/\sqrt{N}$.

Before proceeding we note that, since our approximation is not
self-consistent, we do not attempt to do a proper renormalization by
constructing the counterterms which absorb the divergences of the
integrals. It was shown in \cite{marko10} that even when a strict
expansion in $1/\sqrt{N}$ is performed in the pion propagator
equation, due to the fact that the fermion propagator is unresummed,
different subseries of the counterterms are needed to cancel the
subdivergences of different equations. Here we retain the finite parts
of the integrals obtained using cutoff regularization. 
We refer the interested reader to \cite{marko10} for details.

\subsection{Parametrization\label{ss:param}}

We need to determine the mass parameter $m^2,$ the couplings
$g,\lambda$ of the Lagrangian (\ref{Eq:Lagrange_field}), the
renormalization scale $M_{0B},$ the vacuum expectation value
$v_0\equiv v(T=0,\mu_q=0)$,
and the external field $h,$ which vanishes in the chiral limit. This
is done at $T=\mu_q=0$ using in addition to the pion decay constant
$f_\pi=93$~MeV, pion mass $m_\pi=140$~MeV, and the constituent quark
mass taken to be $M_q=m_N/3=313$~MeV, some information coming from the
sigma sector, such as the mass and the width of the sigma particle and
the behavior of the spectral function.

$v_0$ is determined from the matrix element of the axial vector
current between the vacuum state and a one-pion state, which due to
the rescaling of the vacuum expectation value by $\sqrt{N}$ gives
$v_0=f_\pi/2.$ The value of the Yukawa coupling $g=6.7$ is obtained by
equating the tree-level fermion mass $m_q=g v_0$ with $M_q.$ The
parameters $\lambda$ and $M_{0B}$ are determined from the sigma
propagator, as will be detailed below. Having determined them, in the
chiral limit $m^2$ is fixed from the field equation of $v$, and in
the case of the physical pion mass $m^2$ is determined from the gap
equation by requiring $M^2=m_\pi^2,$ and $h$ is obtained from the
field equation for $v,$ when the local approximation of the pion
propagator is used, while when the approximation (\ref{Eq:Gp_third})
for $G_\pi$ is used, $h$ is fixed by requiring $h=m_\pi^2 v_0,$ and $m^2$
is determined from the field equation of $v$.

In order to fix $\lambda$ and $M_{0B},$ one uses in
(\ref{Eq:Gs}) the tree-level fermion propagator, the field equation
for $v$ (\ref{Eq:EoS}), and the local approximation
(\ref{Eq:Gp_local}) for the pion propagator, which means that 
in the case of a physical pion mass we neglect for simplicity 
the second term of (\ref{Eq:EoS_third}).  After working out the Dirac 
structure one obtains the following form for the sigma propagator:
\bea 
i G_\sigma^{-1}(p)&=&p^2-\frac{h}{v}-\frac{\lambda v^2}{3}
\frac{1}{1-\lambda I_F(p;M)/6}
\nonumber\\
&+&\frac{2 g^2 N_c}{\sqrt{N}} (4m_q^2-p^2)
\tilde I_F(p;m_q).
\label{Eq:Gs_param}
\eea
The integral $I_F(p;M),$ obtained using the local approximation 
(\ref{Eq:Gp_local}) for the pion propagator with $M^2=m_\pi^2,$
 can be found in Eqs.~(10) and (11) of \cite{patkos02} with 
$M_0$ replaced by $M_{0B},$ while $\tilde I_F(p;m_q)$ is given in 
Eqs.~(A16)-(A18) of \cite{marko10}.

\begin{figure}[t]
\begin{center}
\includegraphics[keepaspectratio,width=1.0\columnwidth,angle=0]{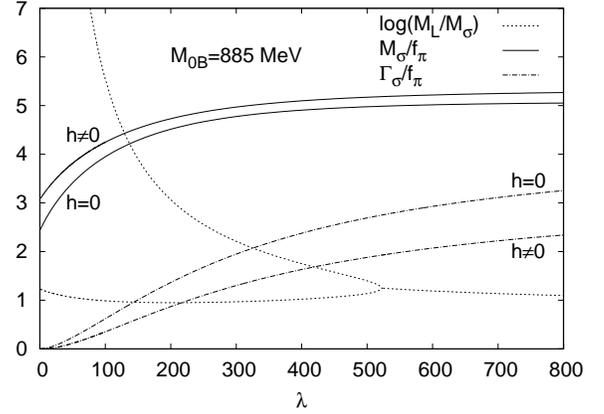}
\caption{The $\lambda$ dependence of the real and imaginary parts of the
complex sigma pole $p_0=M_\sigma-i\Gamma_\sigma/2$ and of the Landau ghost 
$M_L$ in the chiral limit and for the physical pion mass indicated with
label $h\ne 0$ on the curves. $M_L$ is shown only in this latter case, 
for in the chiral limit there is very little difference.}
\label{Fig:sigma}
\end{center}
\end{figure}

The self-energy has both in the chiral limit $M=0$ and for $M=m_\pi$
two cuts along the positive real axis of the complex $p_0$ plane.
These are above the thresholds of the pion and fermion bubble
integrals, which start at $p^2=4 M^2$ and $p^2=4 m_q^2,$
respectively. Above these thresholds the respective pion and fermion
bubble integrals have nonvanishing imaginary parts.  We search for
poles of the sigma propagator analytically continued between the two
cuts to the second Riemann sheet in the form $i
G^{-1}_\sigma(p_0=\kappa e^{-i\phi},\p=0)=0$.  The pole is
parametrized as $p_0=M_\sigma-i \Gamma_\sigma/2,$ with the real and
imaginary parts corresponding to the mass and the half-width of the
sigma particle.  

The solution for $M_\sigma$ and $\Gamma_\sigma$ is
shown in Fig.~\ref{Fig:sigma} both in the chiral limit ($h=m_\pi=0$)
and for the $h\ne 0$ case.  Similar to the case of the $O(N)$ model
studied in Ref.~\cite{patkos02}, in the chiral limit the value of
$M_\sigma$ is a little smaller and the value of $\Gamma_\sigma$ larger
than in the $h\ne 0$ case.  Comparing Fig.~\ref{Fig:sigma} with Fig.~2
of Ref.~\cite{patkos02} obtained in the $O(N)$ model, that is without
fermions, the $M_\sigma(\lambda)$ curve moved slightly upward, but the
$\Gamma_\sigma(\lambda)$ curve moved significantly downward, which
means that in the present case the phenomenologically expected value
\cite{caprini06} $M_\sigma/\Gamma_\sigma\sim 1$ cannot be achieved for
any value of the coupling $\lambda.$ Another difference is that for
low values of $\lambda$ there are two poles of $G_\sigma$ on the
negative imaginary axis in contrast to only one such pole in the
$O(N)$ model. These poles approach each other as $\lambda$ increases
and after they collide at a given value of $\lambda$ there are two
complex poles at higher $\lambda$, one with positive and one with
negative real part. The imaginary part of the complex pole having
positive real part is shown in Fig.~\ref{Fig:sigma} for the
renormalization scale $M_{0B}=885$~MeV. As explained in the study done
in the chiral limit in \cite{toni04} for lower values of the
renormalization scale the scale $M_L$ of the lower Landau ghost on the
imaginary axis comes even closer to $M_\sigma$ and as a result the
spectral function of the sigma is heavily distorted. In order to avoid
this and based on the ratio of $M_\sigma/\Gamma_\sigma$ we have chosen
$\lambda=400$ and $M_{0B}=885$~MeV. For these values
$M_\sigma=456$~MeV and $\Gamma_\sigma=221$~MeV in the chiral case,
while $M_\sigma=474$~MeV and $\Gamma_\sigma=152$~MeV for the case of a
physical pion mass.

\section{The $\bm{\mu_q-T}$ Phase diagram}

The thermodynamics is determined by solving the field equations, 
{\it i.e.} (\ref{Eq:EoS}) and the equations giving the
dependence on $T$ and $\mu_q$ of the two real mean fields $\Phi$ and
$\bar\Phi$, which, when the full fermion propagator is replaced by the
tree-level one have the form:
\bea
\nonumber
\frac{d U(\Phi,\bar\Phi)}{d \Phi}
-2 N_c \sqrt{N}  \int_\k
\frac{k^2}{3 E_k} \left(\frac{d \tilde f_\Phi^+(E_k)}{d \Phi}+
\frac{d \tilde f_{\bar\Phi}^-(E_k)}{d \Phi}
\right)\\
+g^2\sqrt{N} N_c \left[ 
2 \left(\tilde T_F^0(m_q) - T_F(M)\right)
\frac{d \tilde T^\beta(m_q)}{d \Phi}+
\frac{d \tilde T_2^{\beta,2}(m_q)}{d \Phi}\right.
\nonumber\\
-\left.M^2\left(\frac{d S^{\beta,1}(M,m_q)}{d \Phi}
+\frac{d S^{\beta,2}(M,m_q)}{d \Phi}\right)
\right]
=0,\qquad 
\label{Eq:dU_dPhi}
\eea
where $E_k=(\k^2+m_q^2)^{\frac{1}{2}}$ and $M$ satisfies the gap
equation (\ref{Eq:gap_p0}) or the relation $M^2=h/v.$ The other
equation is similar to (\ref{Eq:dU_dPhi}), the only difference is that
the derivative is taken with respect to $\bar\Phi.$ The integral in
(\ref{Eq:dU_dPhi}) is the contribution of the fermionic trace-log
integral, while the term proportional with $g^2$ is the contribution
of the quark-pion two-loop integral in (\ref{Eq:Omega_2PI}) given in
Eq.~(A35) of \cite{marko10}. This term is disregarded for simplicity
when solving the field equations for $\Phi$ and $\bar\Phi$, and only
in one case (see the last row of Table~\ref{tab:phys_data}) the
complete equation (\ref{Eq:dU_dPhi}) is solved in order to estimate
the error made by neglecting it in all the other cases.

The tricritical point (TCP) and the critical end point (CEP) are
identified as the points along the chiral phase transition line of the
$\mu_q-T$ phase diagram where a 1st order phase transition turns
with decreasing $\mu_q$ into a 2nd order or crossover transition,
respectively. In case of a crossover, the temperature $T_\chi$ of the
chiral transition is defined as the value where the derivative $d v/d
T$ has a minimum (inflection point of $v(T)$), while the temperature
$T_d$ of the deconfinement transition is obtained as the location of
the maximum in $d\Phi/d T.$ The transition point in the case of a
1st order phase transition is estimated by the inflection point
located between the turning points of the multivalued curve $v(\mu_q)$
obtained for a given constant temperature. Although the precise
definition of the 1st order transition point is given by that value of
the intensive parameter for which the two minima of the effective
potential are degenerate, we adopt the definition based on the
inflection point because we compute only the derivatives of the
effective potential with respect to the fields and propagators.

\subsection{Phase transition in the chiral limit}

\begin{table}[!b]
\centering
\begin{tabular}{c|c||c|c|c}
\hline
$U(\Phi,\bar\Phi)$ & $\ \ T_0\ \ $ & $T_\chi(0)$ &
$T_d(0)$ & $\ (T,\mu_q)_\ts{TCP}\ $ \\ \hline \hline
$-$           & $-$ & 139.0 &  $-$  &  (60.7,277.0) \\ \hline \hline
poly          & 270 & 185.6 & 229.0 & (104.5,261.8) \\ \hline
poly          & 208 & 168.2 & 176.5 &  (96.2,263.4) \\ \hline
log           & 270 & 191.4 & 209.0 & (109.4,261.2) \\ \hline
log           & 208 & 167.6 & 162.4 & (102.6,261.2) \\ \hline
log  & $T_0(\mu_q)$ & 167.9 & 162.8 &  (84.3,266.9) \\ \hline
Fuku          & $-$ & 176.5 & 193.0 &  (99.8,262.2) \\ \hline
\end{tabular}
\caption{
The (pseudo)critical temperature ($T_d$) $T_\chi$ of the
(deconfinement) chiral transition and the at $\mu_q=0,$ and the 
location of the TCP in units of MeV obtained in 
the chiral limit without the Polyakov loop (first row) and with the 
inclusion of the Polyakov loop using various effective potentials 
summarized in Sec.~\ref{ss:PEP}.
}
\label{tab:chiral_data}
\end{table}

In the chiral limit we solve the field equation (\ref{Eq:EoS}) using
the local approximation to the pion propagator (\ref{Eq:Gp_local})
with $M^2=0$ and neglect the term proportional with $g^2$ in
(\ref{Eq:dU_dPhi}). The critical temperature of the chiral transition
$T_\chi$ and the pseudocritical temperature $T_d$ of the deconfinement
transition at vanishing chemical potential, and the location of the
TCP are summarized in Table~\ref{tab:chiral_data} for various forms of
the Polyakov-loop po\-ten\-tial. With the inclusion of the Polya\-kov
loop $T_\chi(\mu_q=0)$ and $T_\ts{TCP}$ increase significantly
compared with the values obtained earlier in \cite{toni04} without the
Polya\-kov loop, but it has little effect on the value of
$\mu_q^\ts{TCP}.$ This increase in the value of $T_\chi(\mu_q=0)$ is
basically determined by the value of the parameter $T_0$ of the
Polyakov loop potential, while the value of $T_\ts{TCP}$ shows no
significant variation among different cases having the same value of
$T_0$. One can also see, that as explained in \cite{fukushima08}, the
use of the polynomial and logarithmic effective potentials for the
Polyakov loop, that is (\ref{Eq:P_eff_pot_poly}) and
(\ref{Eq:P_eff_pot_log}), drags the value of $T_\chi(\mu_q=0)$ closer
to the value of the parameter $T_0$ than the use of
$U_\ts{Fuku}(\Phi,\bar\Phi)$ given in (\ref{Eq:P_eff_pot_Fuku}). In
this latter case one obtains the smallest value for $T_\ts{TCP}.$

\begin{figure}[!t]
\begin{center}
\includegraphics[keepaspectratio,width=1.0\columnwidth,angle=0]{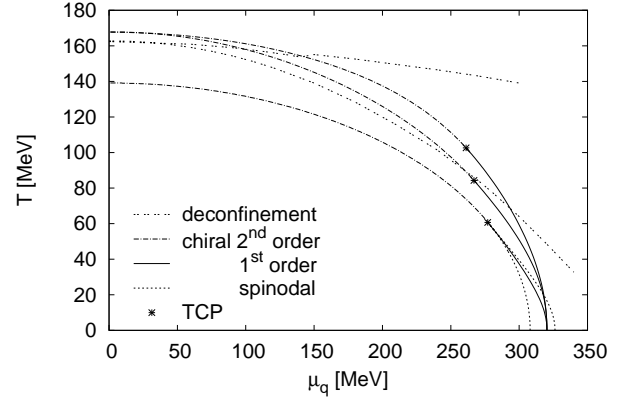}
\caption{
Phase diagrams obtained in the chiral limit without and 
with the inclusion of the Polyakov loop. The former has lower $T_\ts{TCP}$ 
and for the latter we used $U_\ts{log}(\Phi,\bar\Phi)$ with $T_0=208$~MeV 
(upper curves) and with $T_0(\mu_q)$ (middle curves). The deconfinement 
transition line is obtained from the inflection point of $\Phi(T).$
}
\label{Fig:chiral_PD}
\end{center}
\end{figure}

For $T_0=270$~MeV the deconfinement transition line in the $\mu_q-T$
phase diagram is above the chiral transition line in all three
variants of the effective potential for the Polyakov loop.  When the
logarithmic effective potential $U_\ts{log}(\Phi,\bar\Phi)$ is used
either with a constant $T_0=208$~MeV or with the $\mu_q$-dependent
$T_0$ proposed in \cite{schaefer07} one finds $T_d<T_\chi$ at
$\mu_q=0,$ but at a given value of the chemical potential the
deconfinement transition line crosses the chiral transition line and
remains above it for higher values of $\mu_q.$ This is shown in
Fig.~\ref{Fig:chiral_PD}, where the deconfinement transition line is
obtained from the inflection point of $\Phi(T).$ In contrast to the
case of constant $T_0,$ where basically the deconfinement transition
line is not affected by the increase of $\mu_q,$ with a
$\mu_q$-dependent $T_0$ the deconfinement transition line strongly
bends, staying close to the chiral line. The two lines cross just
above the TCP.

In the case when $T_0(\mu_q)$ is used, the lowering of the
deconfinement transition results in the shrinking of the region of the
$\mu_q-T$ plane for which $T_\chi<T<T_d,$ already observed in
Ref.~\cite{abuki08}. Since the quantity measuring the quark content
inside thermally excited particles carrying baryon number shows a
pronounced change along the chiral phase transition line of this
region, the region was identified in \cite{fukushima08} with the
so-called quarkyonic phase, a confining state made of quarks and is
characterized by a high quark number density and baryonic (three-quark
state) thermal excitations.

Comparing our results on the phase diagram to those obtained in the
chiral limit of the PNJL model one can notice differences of both
qualitative and quantitative nature. In the nonlocal PNJL model of
Ref.~\cite{sasaki07} the deconfinement phase transition line starts at
$\mu_q=0$ below the chiral transition line both for a polynomial and a 
logarithmic Polyakov-loop effective potential with $T_0=270$~MeV, so
that the two transition lines cross at finite $\mu_q.$ In our case
this happens only for the logarithmic potential with $T_0=208$~MeV, as
can be seen in Fig.~\ref{Fig:chiral_PD}. In \cite{sasaki07,costa09a}
the values of $T_\chi(\mu_q=0)$ and $T_\ts{TCP}$ are much larger than
in our case, while the value of $\mu_q^\ts{TCP}$ is similar to ours.

\subsection{Phase transition with a physical pion mass}

\begin{table}[b]
\centering
\begin{tabular}{c|c||c|c|c|c}
\hline
$U(\Phi,\bar\Phi)$ & $\ T_0\ $ & $T_\chi(0)$ & $T_d(0)$ &
$\Gamma_\chi$ & $\ (T,\mu_q)_\ts{CEP}\ $ \\ \hline \hline
$-$  & $-$ & 158.6 &  $-$  & 40.7 & (13.5,328.6) \\\hline\hline
poly & 270 & 212.5 & 217.4 & 28.3 & (32.9,328.8) \\ \hline
poly & 208 & 184.6 & 176.8 & 22.3 & (30.6,328.8) \\ \hline
log  & 270 & 209.7 & 209.3 & 12.0 & (34.5,329.0) \\ \hline
log  & 208 & 168.5 & 167.1 & *43.0& (33.0,328.9) \\ \hline
Fuku & $-$ & 195.2 & 191.3 & 21.2 & (31.8,328.8) \\ \hline \hline
poly & 208 & 188.1 & 183.1 & 21.4 & (32.2,329.0) \\ \hline
\end{tabular}
\caption{
The temperatures $T_\chi$ and $T_d$ of the chiral and deconfinement
transitions, the half-width at half maximum $\Gamma_\chi$ of $-d v/d
T$ at $\mu_q=0$ (in the case marked with $*$, due to the asymmetric
shape of $-d v/d T,$ the full width is given) and the location of the
CEP in units of MeV obtained using (\ref{Eq:Gp_third}) for the pion
propagator without and with the inclusion of the Polyakov loop.  The
contribution of the quark-pion setting-sun was kept in
(\ref{Eq:dU_dPhi}) only for the result of the last row.}
\label{tab:phys_data}
\end{table}

In the approximation (\ref{Eq:Gp_third}) for the pion propagator,
which resum infinitely many orders in $1/\sqrt{N},$ the phase
transition at $T=0$ turns with increasing $\mu_q$ from a crossover
type into a first order transition at some value $\mu_q^c>M_q,$ and in
consequence there is a CEP in the $\mu_q-T$ phase diagram. The
numerical results are summarized in Table~\ref{tab:phys_data} for
various forms of the Polyakov-loop potential reviewed in
Sec.~\ref{ss:PEP}.  Increasing the temperature $\mu_q^c$ decreases and
the first order chiral restoration becomes a crossover at a much lower
temperature $T_\ts{CEP}$ than in the chiral case. The inclusion of the
Polyakov loop increases significantly the value of $T_\ts{CEP},$ but,
as in the chiral case, it has little effect on the value of
$\mu_q^\ts{CEP}.$ Neither the choice of the effective potential for
the Polyakov loop nor the value of $T_0$ has a significant effect on
the value of $\mu_q^\ts{CEP}.$ The result in the last row was obtained
by keeping in the field equation of the Polyakov loop
(\ref{Eq:dU_dPhi}) and its conjugate the contribution of the
quark-pion setting-sun diagram, while in all other cases only the
contribution of the fermionic trace-log was kept. Comparing the result
in the last row of Table~\ref{tab:phys_data} with that of the second
row obtained using the polynomial Polyakov-loop potential, one sees
that the error we make by neglecting the setting-sun contribution in
all other cases is fairly small.

\begin{figure}[!t]
\centering
\includegraphics[keepaspectratio,width=0.495\textwidth,angle=0]{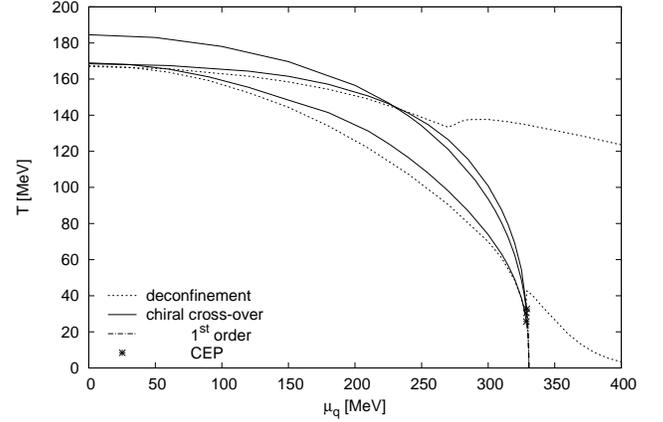}
\caption{
Phase diagrams obtained for the physical value of the pion mass with the
inclusion of the Polyakov loop.  For the chiral transition line which 
starts at higher $T$ for $\mu_q=0$ we used $U_\ts{poly}(\Phi,\bar\Phi)$ with
$T_0=208$~MeV and (\ref{Eq:Gp_third}) for the pion propagator, for the other
two phase diagrams we used the local approximation for the pion propagator
and $U_\ts{log}(\Phi,\bar\Phi)$ with $T_0=208$~MeV (middle curves) and with
$T_0(\mu_q)$ (lower curves).  The deconfinement transition line is obtained
from the inflection point of $\Phi(T).$
}
\label{Fig:phys_PD} 
\end{figure}

The values of $T_\chi$ and $T_d$ at $\mu_q=0$ are mostly influenced by
the choice of the Polyakov effective potential and the value of $T_0:$
they decrease with the decrease of $T_0$ and by using the logarithmic
potential instead of the polynomial one. Using the polynomial
potential with $T_0=270$~MeV the confinement transition line in the
$\mu_q-T$ plane is above the chiral transition line. As in the chiral
case, when a logarithmic potential is used with either a fixed value
$T_0=208$~MeV or with a $\mu_q$-dependent $T_0,$ the deconfinement
transition line starts at $\mu_q=0$ below the chiral one and the two
lines cross at some higher value of $\mu_q.$ This can be seen in 
Fig.~\ref{Fig:phys_PD}. When $T_0(\mu_q)$ is used the
two lines go together until they cross each other just above the
location of the CEP. This $\mu_q$-dependent $T_0$ gives the lowest
value of $T_\ts{CEP},$ similar to the results reported in
\cite{ciminale08} and \cite{herbst10}. Because of the much lower value
of the $T_\ts{CEP}$ the shrinking of the quarkyonic phase is more
pronounced than in the chiral case, as the deconfinement transition
lines approaches the $\mu_q$ axis.  This is even more the case here,
with a physical pion mass, since the deconfinement transition is a
crossover and as such it happens in a relatively large temperature
interval. However, the quarkyonic phase does not vanish completely as
happens in \cite{herbst10}, where quantum fluctuations are included
using functional renormalization group methods.

\section{Conclusions}

Using the tree-level fermion propagator and some approximations for the
self-consistent pion propagator obtained within a large-$N_f$ expansion, we
studied in the $SU(2)_L\times SU(2)_R$ chiral quark-meson model, in the
chiral limit and for the physical value of the pion mass, the influence of
the Polya\-kov loop on the chiral phase transition.  When the local part of
the approximate pion propagator resums infinitely many orders in $1/N_f$ of
fermionic contributions it is possible to find a CEP on the chiral phase
transition line of the $\mu_q-T$ phase diagram.  The inclusion of the
Polyakov loop potential has a significant effect on $T_\ts{CEP}$ and
practically no effect on $\mu_q^\ts{CEP}$ obtained in the original chiral
quark-meson model, that is which does not contain the Polyakov loop.  Using
the logarithmic form $U_\ts{log}(\Phi,\bar\Phi)$ of the effective potential
for the Polyakov loop with parameter $T_0=208$~MeV a crossing between the
chiral and deconfinement transition lines was observed, with the latter line
starting at $\mu_q=0$ slightly below the former one.  In this case the
existence of the quarkyonic phase is possible.

It was shown in \cite{marko10} that the result of resumming in the
pion propagator ${\cal O}(1/\sqrt{N})$ fermionic fluctuations obtained
with a strict expansion in $1/\sqrt{N}$, while keeping the fermion
propagator unresummed, the phase transition softens to the point that
there is no CEP in the $\mu_q-T$ phase diagram within a range
$0<\mu_q<500$~MeV. For this reason it is an interesting question to
what extent our results in the existence and location of the CEP would
be modified by the use of the self-consistent propagator for fermions,
and also by considering the more realistic $SU(3)_L\times SU(3)_R$ 
chiral quark-meson model.

\section*{Acknowledgments} 
This work is supported by the Hungarian Research Fund
under Contracts No.~T068108 and No.~K77534.

\end{document}